\documentclass[aps,prb,twocolumn,superscriptaddress]{revtex4}
\usepackage{graphicx}
\usepackage{feynmp}
\usepackage{epsfig}
\usepackage{amsmath}
\usepackage{color}

\newcommand{\lder}[1]{#1 \frac{\partial}{\partial#1}}

\font\elevenmib=cmmib10 scaled 1095
\font\tenmib=cmmib10
\font\eightmib=cmmib10 scaled 800
\font\sixmib=cmmib10 scaled 667
\skewchar\elevenmib='177
\newfam\mibfam

\textfont\mibfam=\tenmib
\scriptfont\mibfam=\eightmib
\scriptscriptfont\mibfam=\sixmib

\mathchardef\sigma="711B

\newcommand{\beq}{\begin{equation}}
\newcommand{\eeq}{\end{equation}}

\def\bK{\mathbf{K}}

\def\bP{\mathbf{P}}

\begin{document}

\title{The Renormalization Group and the Superconducting Susceptibility of a Fermi Liquid}


\author{S. A. Parameswaran} 
\affiliation{Department of Physics, Joseph Henry Laboratories, Princeton University, Princeton, New Jersey 08544, USA}
\author{R. Shankar}
\affiliation{Department of Physics,  Yale University, New Haven, CT 06520, USA}
\author{S. L. Sondhi}
\affiliation{Department of Physics, Joseph Henry Laboratories, Princeton University, Princeton, New Jersey 08544, USA}

\date{\today}

\begin{abstract}

A free Fermi gas has, famously, a superconducting susceptibility that diverges logarithmically at zero temperature. In this paper we ask whether this is still true for a Fermi liquid and find that the answer is that it does {\it not}. From the perspective of the renormalization group for interacting fermions, the question arises because a repulsive interaction in the Cooper channel
is a marginally irrelevant operator at the Fermi liquid fixed point and thus is also expected to infect various physical
quantities with logarithms. Somewhat surprisingly, at least from the renormalization group viewpoint, the result for the superconducting susceptibility is that two logarithms are not better than one. In the course of this investigation we
derive a Callan-Symanzik equation for the repulsive Fermi liquid using the momentum-shell renormalization group, and use it to compute the long-wavelength behavior of the superconducting correlation function in the emergent low-energy theory. We
expect this technique to be of broader interest.

\end{abstract}

\maketitle
\section{Introduction}
Thinking within the framework of the renormalization group (RG) is always an insightful way to approach the low energy, long wavelength behavior of physical systems, but it does not always provide a significant computational advantage over more
straightforward, perturbative, methods. The exceptions are cases that involve marginal and near marginal couplings where
making sense of perturbative divergences is far easier within the framework of the RG. Indeed, the historical development
of RG methods had mostly to do with exactly these cases leading up to the discovery of the epsilon expansion by Wilson and Fisher \cite{Wilson1974,Fisher1974}.

A subclass of these canonical RG applications consists of problems with marginally relevant or irrelevant couplings. Celebrated examples of the former are the Kondo problem\cite{AndersonYuvalHamann1970,WilsonKondo1975} and
QCD\cite{GrossWilczek1973} while the textbook example of the latter is the critical behavior of
ferromagnetic models (equivalently, of vector models in the language of field theory) in four dimensions\cite{ZinnJustin}. In all of these
cases, non-trivial logarithms appear in physical quantities either in the ultraviolet (marginally relevant) or the infrared (marginally irrelevant), and reflect the extremely slow variation of the coupling at issue. An instructive example, with
close parallels to our concerns in this paper, is the critical behavior of the specific heat in four dimensional $O(N)$
ferromagnets/vector models. Here the non-interacting gaussian theory exhibits a logarithmic divergence
$$
C \sim -\log |t| \ \ {\rm with} \ \ t = \frac{T-T_c}{T_c} \ .
$$
As the quartic coupling is marginally irrelevant, i.e., it dies logarithmically in the infrared and the theory flows to the non-interacting limit, one may then suppose that  the divergent susceptibility of free-field theory will remain intact even upon inclusion of the interactions. However, as is long established, interactions lead to the more complex logarithmic dependence
$$
C \sim [-\log |t|]^{\frac{4-N}{N+8}}
$$
which {\it diverges} as $(- \log |t|)^{1/3}$ in the Ising case (N = 1) but {\it vanishes} as $(-\log |t|)^{-1}$ when $N \rightarrow \infty$.
Observe that the logarithms appear generically with fractional powers and a variable sign. This complexity is not easily unraveled perturbatively and instead it is much better to resort to the machinery of the RG, specifically to the derivation and
solution of the RG (differential) equations obeyed by the correlation functions, which we shall simply call Callan-Symanzik equations in this paper.

Let us turn now to the problem considered in this paper. The RG treatment of interacting fermions identifies the Landau Fermi liquid as a fixed point of a momentum shell RG characterized by set of exactly marginal couplings consisting of the Landau $F$
function. In addition however, there is a marginally irrelevant coupling which is the repulsive BCS (Cooper channel)
coupling.\footnote{If the BCS coupling is attractive, it is marginally relevant and we have the physics of the superconducting instability.} Absent this marginal flow the fermions have a logarithmically divergent susceptibility to superconductivity. We
wish to ask what happens to this divergence when the marginally irrelevant flow is taken into account. We show here using the RG that this divergence goes away and the uniform, zero frequency superconducting susceptibilty of the Fermi liquid is therefore
{\it finite}.

It is useful at this point to clarify what we mean by the Fermi liquid. As the reader is no doubt aware, the actual RG flow for an interacting fermion system
with typical repulsive interactions leads inevitably to a superconducting instability via the Kohn-Luttinger effect wherein screening produces an effective
interaction that is attractive in a higher angular momentum channel, as discussed for example in Ref~\onlinecite{RSRMP}. What we have in mind therefore is
the RG flow for a system where the dominant bare couplings in the BCS channel are repulsive and renormalize to smaller values for a large range
of scales (temperature or energy) while the growing attractive couplings are still small---operationally this is what one means by a Fermi liquid. We
can, however, formalize this understanding by working with Hamiltonians which contain {\it only} the Landau couplings and the reduced BCS
couplings that are left in a naive application of the RG to the interacting fermion problem. In this approximation, the Kohn-Luttinger scale has
truly vanished but we are still left with a marginally irrelevant operator about the Fermi liquid Hamiltonian whose flow can be studied.

Our central result is possibly not new---certainly it is almost present in the large lore on superconductivity and
superconducting fluctuations\cite{VarlamovLarkin} and we would be delighted to hear from readers who can point us to an
 explicit, relevant citation. It does not, however, appear to be widely known and at first blush comes as a surprise, as condensed matter theorists are
conditioned to think of the Fermi liquid as exhibiting correlations morally identical to that of the Fermi gas up to
the effects of the Landau parameters. {\it Ex post facto} the intuition we would offer is that the repulsive interaction
goes away just slowly enough at long distances that the usual buildup of the divergence of the superconducting susceptibility
is undone. Possibly there is a deformation of our present problem where this exact cancellation can be modified to yield
a more complex residual of the kind cited for four dimensional ferromagnetism above.

In any event, we view our work as a contribution to the RG analysis of interacting fermions and as such we trust readers will
find it interesting as well. In the process, we show how to derive Callan-Symanzik equations to one loop for interacting fermions for composite operators made from the fundamental Fermi fields---which may be of interest to readers even beyond the specifics
of our application to the computation of the superconducting susceptibility.

In the following we begin with a quick summary, in Section II, of the RG for interacting fermions that leads to the Fermi liquid fixed point(s) and the flow in its vicinity.  Next we derive the relevant Callan-Symanzik equations (Section III), solve them
(Section IV) and end with a brief discussion (Sections V and VI) and an Appendix that contains some technical details. Note that while our results certainly apply in both two and three dimensions, to avoid unnecessarily complications we work with the pedagogically simpler case of $d=2$ throughout this paper.

\section{Review of Results from the RG}

We begin the technical part of our discussion with a summary of results from the renormalization group approach as applied to interacting fermion systems. We shall provide a telegraphic review, referring the reader interested in further details  to more pedagogical discussions, such as  Ref.~\onlinecite{RSRMP}. Readers familiar with the technology and results can skip ahead to Section III.

We  focus on the  following following action, written for a system of electrons with a circular Fermi surface in $d=2$ and spin directions $\alpha  =\uparrow,\downarrow$ :
\begin{widetext}\begin{eqnarray}
S &=&\sum_{\alpha=\uparrow,\downarrow}\int_{-\infty}^{\infty}\frac{d\omega}{2\pi}\int_0^{2\pi}\frac{d \theta}{2\pi} \int_{-\Lambda}^{\Lambda} \frac{dk}{2\pi} \, \bar{\psi}_\alpha(\omega\theta k) \left(i\omega -v_Fk \right)\psi_\alpha(\omega\theta k)+  \sum_{\begin{subarray}{c} \mu,\nu\\\alpha\beta\end{subarray}}\,\int_{\substack{\{k_i\}\\ \{\omega_i\}}} u_{\mu\nu\alpha\beta}(1,2,3,4) \bar{\psi}_\mu(1) \bar{\psi}_\nu(2)\psi_\alpha(3)\psi_\beta(4)\nonumber\\
&\equiv & S_0 + S_I
\end{eqnarray}\end{widetext}
where $\psi_\mu(i) = \psi_\mu(\omega_i\theta_i k_i)$ is the Fermion/Grassman field, $k = |\mathbf{K}|-K_F$ is the radial
component of  momentum measured relative to the Fermi momentum $K_F$, $v_F$ is the Fermi velocity, $\Lambda <<K_F $ is the ultraviolet cutoff.

In the second term $S_I$, the measure is \begin{equation} \int_{\substack{\{k_i\}\\ \{\omega_i\}}}= \prod_{\begin{subarray}{c}i=1\\ \mathbf{K}_4=\mathbf{K}_1+\mathbf{K}_2-\mathbf{K}_3\end{subarray}}^3\int_{-\infty}^{\infty} d\omega_i \int_0^{2\pi}\frac{d\theta_i}{2\pi}\int_{-\Lambda}^{\Lambda} \frac{dk_i}{2\pi} \Theta(\Lambda-|\mathbf{k}_4|).\label{measure} \end{equation}

The renormalization group transformation involves  three steps: (i) integrating out all momenta between $\Lambda/s$ and $\Lambda$, and correcting terms in the action as needed in the process; (ii) rescaling frequencies  and the momenta  as per $(\omega , k ) \rightarrow s\ (\omega, k$ ) so that the  cutoff in $k$ is once again at $\pm \Lambda$; and finally (iii) rescaling fields $\psi \rightarrow Z_{\psi}\psi$ to keep the free-field action $S_0$  invariant. Thus $S_0$ is a fixed point of this RG and the possible interactions can be classified as relevant, irrelevant, or marginal with respect to this transformation and fixed point.

At tree level, it is easily shown that interactions with six or more fields are irrelevant, as is  the $\omega$ or $k$ dependence of quartic couplings. While this is very much like $\phi^{4}_{4}$, the $\phi^4$ theory in four dimensions, there the coupling is just a {\em number}, $u(0,0,0,0)$,  describing collision of particles at zero external momentum, here the quartic couplings can  depend on the angles on the Fermi circle. Given that momentum is conserved, we need to pick only three of the angles independently, say $\theta_1,\theta_2$ and $\theta_3$. However, the fact that the momenta come not from the plane but a very thin annulus leads to additional constraints \cite{RSRMP}.

\begin{figure}
\includegraphics[width=8cm]{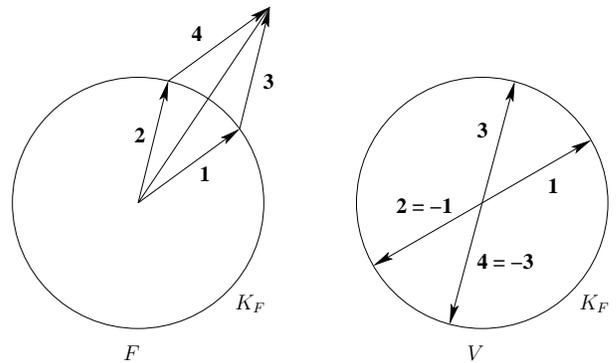}
\caption{\label{kinematics}
Kinematics of the couplings F and V.}
\end{figure}

Consider the left half of Figure \ref{kinematics} where  all momenta lie on the Fermi circle. Given $1$ and $2$ , it is clear that $3$ and $4$ have to equal them pairwise. That is $\theta_1=\theta_3$ and $\theta_2=\theta_4$ or the exchanged version. Now consider the generic case of   momenta that lie in a very thin shell $|k|<\Lambda$, rather than right on the Fermi circle. It is not surprising  that now $\theta_1\simeq \theta_3$ and $\theta_2\simeq \theta_4$ with deviations of order $\Lambda /K_F$. We ignore the
dependence of $u$ on such tiny angular differences when $\Lambda /K_F \to 0$, and we ignore any $k$ -dependence (since it is irrelevant). Thus we define $u$ with all $k=0$ and $\theta_1=\theta_3$ and $\theta_2=\theta_4$:
\begin{eqnarray}
u (\theta_1, \theta_2,\theta_3, \theta_4)&=& u (\theta_1, \theta_2,-\theta_1, -\theta_2) =u(\theta_1, \theta_2)\nonumber \\
&=& u(\theta_1- \theta_2)\equiv F(\theta).
\end{eqnarray}

It is important to bear in mind that {\em even though we evaluate  $u$ at $k=0$  for angles $(\theta_1= \theta_3)$ and $(\theta_2=\theta_4)$,}  we do not imply that only forward scattering is allowed or that all momenta lie on the Fermi surface:  rather we allow all momenta in the measure defined by Eqn. \ref{measure} but ignore the dependence of $u$ on $k$ and the tiny differences  $(\theta_1- \theta_3)$ and $(\theta_2-\theta_4)$. (It is like saying that in $\phi_{4}^{4}$, $u=u(0,0,0,0)$ does not mean the external legs are  limited to zero, only that $u$ is the same for all values of external momenta.) Indeed a small amount of non-forward scattering is not only allowed at any nonzero cut-off, it is essential  to  produce a nonzero compressibility in the ``$q$" limit.

In summary, in studying the flow of the four point coupling, we will  choose the legs to be on the Fermi surface (since $k$ dependence is irrelevant), i.e, we study the flow of  $F$.

Kinematics allows one more coupling function besides $F$. Consider the right half of Figure \ref{kinematics} with all momenta on the Fermi circle,   but with $\theta_1 = -\theta_2$ so that  the incoming momenta add up to zero. Now  the outgoing pair of momenta can point in {\em any} direction, as long as they add up to zero, which lets them point in any pair of mutually opposite directions. This leads to a function $V$ in the BCS channel:
\begin{eqnarray}
u (\theta_1, \theta_2,\theta_3, \theta_4)&=& u (\theta_1, -\theta_1,\theta_3, -\theta_3) =u(\theta_1, \theta_3)\nonumber \\ &=& u(\theta_1- \theta_3)\equiv V(\theta)
\end{eqnarray}
Once again, even though  $u$ will be assumed to depend on just the directions $\theta_1$ and $\theta_3$ (via their difference), for any $\Lambda >0$, scattering  between states of the total momentum  $P\simeq \Lambda$ will be kinematically permitted.

The fate of the couplings $F$ and $V$, marginal at tree level, is determined by one loop diagrams.
The one-loop analysis of Ref~\onlinecite{RSRMP} shows that $F$ is strictly marginal while $V$ flows. It is also shown there that the one-loop flow is exact in the limit $\Lambda /K_F \to 0$.

In the remainder of this paper, we will assume  $F=0$  and that  $V$  is repulsive and angle independent. While this simplifies the discussion, the first and last restrictions are inessential. Inclusion of $F$ will not affect the physics of the Cooper channel for kinematic reasons, and while the angular dependence of $V$ can affect details of various calculations, it cannot alter the fact that singular logarithms arise in the same.  As long as the system remains rotationally invariant,  any angular dependence in $V$ can be accounted for by expanding in different angular momentum channels, and deriving flow equations for each channel. Each channel will have its own Callan-Symanzik equation, and the remainder of the arguments follow from this. On the other hand,  it should be clear from the preceding discussion that restricting $V$ to be repulsive is  necessary, since attraction leads to the BCS instability, and of course the Fermi liquid fixed point cannot capture the physics of superconductivity. 

Thus,
\begin{equation}
S_I= \sum_{\begin{subarray}{c} \mu,\nu\\\alpha\beta\end{subarray}}\,\int_{\substack{\{k_i\}\\ \{\omega_i\}}} \frac{g_0}{4}  \epsilon_{\mu\nu}\epsilon_{\alpha\beta}  \bar{\psi}_\mu(1) \bar{\psi}_\nu(2)\psi_\alpha(3)\psi_\beta(4)\Theta (\lambda -|P|)
\end{equation}
where the additional condition $ \Theta (\lambda -|P|) $ ensures that the total momentum of the Cooper pair is less than $\lambda$ which itself is assumed to obey $\lambda <<\Lambda$. This will not be a limitation in our study of superconductivity since only the interaction at $\bP\simeq 0$ will come into play.

We will use the  standard parametrization of $s$:

\beq
s= e^{\sigma}
\ \ \mbox{so that} \ \ \    \ d\sigma ={ds \over s}= -{d\Lambda \over \Lambda}.
\eeq
We will typically choose to rescale by an amount close to unity, in which case
\beq
 d\sigma  = {ds \over s}  = ds \ \ \   \mbox {since  } \ \  s \to 1.
 \eeq
Note that the cutoff is being changed infinitesimally at each RG step; as the RG transformations are infinitesimal, this permits us to derive differential equations that capture the content of the flows.

It is shown in Ref. ~\onlinecite{RSRMP} that
\beq
\beta(g) \equiv {dg \over d\sigma} =  -ag^2
\eeq
where $a$ is a positive constant. The solution to this equation  is
\beq
g(\sigma ) = \frac{g_0}{ 1+ a\sigma  g_0}.
\eeq
For repulsive $g_0$ we see a logarithmically vanishing $g(\sigma ) \simeq 1/\sigma$. One may expect that such a marginally irrelevant coupling cannot affect the superconducting susceptibility which diverges logarithmically in the noninteracting limit. The rest of this paper aims to show why this is not the case and what finally happens.

\section{Derivation of the Callan-Symanzik Equations}
So far,  we have reviewed  how different coupling functions in the Hamiltonian flow under the RG. This allowed us to classify relevant and irrelevant perturbations, and determine that the repulsive Cooper channel interaction, being marginally irrelevant, does not lead to an instability: the system flows.  In this section, we discuss how to derive a set of differential equations that describe how the correlation functions evolve under the RG flow. First, we discuss how to derive the equations for correlation functions without composite operators, and then discuss the procedure for the Cooper pair operator.

\subsection{Callan-Symanzik Equations for fundamental Fermi field correlators}
We begin with the  correlation function \footnote{We suppress spin indices for convenience.}, defined with an action $S$ and cutoff $\Lambda$:
\begin{equation}
\langle \bar{\psi}_{\bar\omega\bar\theta \bar k} \psi_{\omega \theta k} \rangle_{S, \Lambda} = \frac{\displaystyle{\int\left[D\bar\psi D\psi\right]_{\Lambda}}e^{-S_{\Lambda}[\bar\psi,\psi]} \bar\psi_{\bar\omega \bar\theta  \bar k}\psi_{\omega \theta k}}{\displaystyle{\int\left[D\bar\psi D\psi\right]_{\Lambda}}e^{-S_{\Lambda}[\bar\psi,\psi]}}
\end{equation}
where $\left[D\bar\psi D\psi\right]_{\Lambda}$ indicates that the functional integral is over modes of all momenta upto $\Lambda$. Here it is understood that the momenta $k$ are well within the cutoff. Note also that the momenta and frequencies corresponding to $\bar\psi$ and $\psi$ are different; in fact, the result will have a delta function that forces these to be the same, but since the delta function is dimensionful, it affects the scaling, so we maintain different momenta explicitly through the calculation. Integrating out a shell $d\Lambda$ of momenta between $\Lambda$ and $\Lambda+d\Lambda =\Lambda - |d\Lambda| \equiv \Lambda/s \equiv \Lambda/(1+d\sigma)$, we have \cite{RSRMP}
\begin{equation}
\langle \bar{\psi}_{\bar\omega\bar\theta \bar k} \psi_{\omega \theta k} \rangle_{S, \Lambda} = \frac{\displaystyle{\int\left[D\bar\psi^{<} D\psi^{<}\right]_{\frac{\Lambda}{s}}}e^{-S^{eff}_{\Lambda/s}[\bar\psi^{<},\psi^{<}]} \bar\psi^{<}_{\bar\omega\bar\theta \bar k}\psi^{<}_{\omega \theta k}}{\displaystyle{\int\left[D\bar\psi^{<} D\psi^{<}\right]_{\frac{\Lambda}{s}}}e^{-S^{eff}_{\Lambda/s}[\bar\psi^{<},\psi^{<}]}}
\end{equation}
where $S^{eff}_{\Lambda/s}$ is an action for a theory with cutoff $\Lambda/s$, with the appropriate corrections to parameters, and the $<$ superscript denotes `slow' modes that lie within the reduced cutoff. In order to obtain a theory with the \emph{same} cutoff, we rescale momenta in the second equation. We will write $\psi'(\omega' \theta' k') = \tilde{Z}_\psi^{-1/2}\psi^{<}(\omega'/s, \theta, k'/s)$, where $0<|k'|<\Lambda$. Further, we assume that action may be written as $S^{eff}_{\Lambda/s}[\bar\psi^{<},\psi^{<}]  = S_{\Lambda}[\bar\psi',\psi'] +\delta S_{\Lambda}[\bar\psi',\psi']$, where $\delta S$ is $O(d\sigma)$. We obtain
\begin{eqnarray}
\langle \bar{\psi}_{\bar\omega\bar\theta\bar k} \psi_{\omega \theta k} \rangle_{S, \Lambda} &=& \tilde{Z}_{\psi} \frac{\displaystyle{\int\left[D\bar\psi' D\psi'\right]_{\Lambda}}e^{-S_{\Lambda} +\delta S_\Lambda} \bar\psi'_{s\bar\omega \bar\theta s \bar k}\psi'_{s\omega \theta sk}}{\displaystyle{\int\left[D\bar\psi' D\psi'\right]_{\Lambda}}e^{-S_{\Lambda}+\delta S_{\Lambda}}}\nonumber\\&=& \tilde{Z}_{\psi}\langle \bar\psi_{s\bar\omega, \bar\theta,s \bar k} \psi_{s\omega,\theta,sk} \rangle_{S+\delta S, \Lambda}
\end{eqnarray}
where in the second step we dropped the primes as the fields are integrated over.

Since $\delta S$ depends on the parameter $s$ of the flow, we can write $S+\delta S = S_{s}$; then, we have
\begin{equation}
\langle \bar{\psi}_{\bar\omega\bar\theta\bar k} \psi_{\omega \theta k}\rangle_{S,\Lambda}  = \tilde{Z}_{\psi}\langle \bar\psi_{s\bar\omega,\bar \theta,s\bar k} \psi_{s\omega,\theta,sk} \rangle_{S_s, \Lambda}
\end{equation}

The left side of this equation is manifestly independent of $s$. Therefore, differentiating both sides with respect to $s$,

and dividing through by $\tilde{Z}_{\psi}$,

\begin{equation}
\frac{d\log\tilde{Z}_{\psi}}{ds}\langle \bar\psi_{s\bar\omega, \bar\theta,s\bar k} \psi_{s\omega,\theta,sk} \rangle_{S_s, \Lambda} + \frac{d}{ds}\langle \bar\psi_{s\bar\omega, \bar\theta,s\bar k} \psi_{s\omega,\theta,sk} \rangle_{S_s, \Lambda} =0\nonumber
\end{equation}

The second term can be rewritten as \footnote{We note that $\mathcal{Z}$, the partition function, is independent of $s$, and implicitly use the fact that the correlation function forces $\omega =\bar\omega$, etc; in a moment, we will switch to a discussion of the Green's function, and this complication will be avoided.}

\begin{eqnarray}
\frac{d}{ds}\langle \bar\psi_{s\bar\omega,\bar\theta,s \bar k} \psi_{s\omega,\theta,sk} \rangle_{S_s, \Lambda} & &\nonumber\\
= \left[ \lder{\omega}+\lder{k} \right.&+&\left.  \frac{dg}{ds} \frac{\partial}{\partial g}\right]  \langle \bar\psi_{s\bar\omega,\bar \theta,s \bar k} \psi_{s\omega,\theta,sk} \rangle_{S_s, \Lambda}\nonumber\\
\end{eqnarray}
where we note that the only change in the action to one loop order is in the coupling constants. Collecting terms, and taking the $s\rightarrow 1$ limit, we arrive at the Callan-Symanzik equation for the two-point correlation function,
\begin{equation}
\left[ \lder{\omega} +\lder{k} +{\beta}(g)\frac{\partial }{\partial g} +3-2{\gamma}_{\psi} \right]\langle \bar{\psi}_{\bar\omega \bar\theta \bar k} \psi_{\omega\theta k}  \rangle_{S, \Lambda} =0
\end{equation}
where  ${\beta}(g) = \left.\frac{dg}{ds}\right|_{s\rightarrow1}= \frac{dg}{d\sigma}$ is the beta function, and  we define the anomalous dimension $3-2{\gamma}_{\psi} = \left.\frac{d\log\tilde{Z}_{\psi}}{ds}\right|_{s\rightarrow 1}$.

Next, we observe that the Green's function is related to the correlation function as $\langle \bar\psi_{\bar\omega \bar\theta \bar k} \psi_{\omega\theta k} \rangle_{S, \Lambda} =  G(\omega,\theta, k) \delta_{\omega,\bar\omega}\delta_{\theta,\bar\theta}\delta_{k,\bar k}$. Since the $\delta$ functions each have dimension $-1$, it is easily verified that the Green's function satisfies the relation

\begin{equation}
\left[ \lder{\omega} +\lder{k} +{\beta}(g)\frac{\partial }{\partial g} +1-2{\gamma}_{\psi} \right] G(\omega, \theta, k) =0
\end{equation}

Finally, we rewrite this as an equation for the amputated Green's function, $\Gamma^{(2,0)}_{\psi} \equiv G^{-1}$:

\begin{equation}
\left[ \lder{\omega} +\lder{k} +{\beta}(g)\frac{\partial }{\partial g} -1+2{\gamma}_{\psi} \right] \Gamma^{(2,0)}_{\psi} (\omega, \theta, k) =0
\end{equation}

We see that for the free theory, where $\beta = \gamma =0$, $ \Gamma^{(2,0)}_{\psi} \sim i\omega -v_Fk$ satisfies the above relation, as expected.

\subsection{The Callan-Symanzik Equation for the Cooper Pair Operator}
The central objects of this paper are the Cooper pair operator and its two-point correlation function. We define the composite operator that creates an $s$-wave Cooper pair at frequency $\Omega$ and momentum $\bP<<\Lambda $ by
\begin{eqnarray}
\bar{\mathcal{O}}_{\Omega,\bP}&=& \left(\bar\psi \bar\psi\right)_{\Omega,\mathbf{P}} \nonumber\\&=& \int_{-\infty}^{\infty}\frac{d\omega}{2\pi}\int_{-\Lambda}^{\Lambda}{dk \over 2\pi} \int_{0}^{2\pi} \frac{d \theta }{2\pi}\, \bar{\psi}_{\omega+\Omega,\mathbf{K}+\frac{\mathbf{P}}{2},\uparrow} \bar\psi_{-\omega, -\mathbf{K} +\frac{\mathbf{P}}{2}, \downarrow}\nonumber\\
\end{eqnarray}
where $(k,\theta )$ refer to $\mathbf{K}$. (Since $P<<\Lambda$, if $\mathbf{K}$ lies in the shell so will $\mathbf{K} +\frac{\mathbf{P}}{2}$.) The operator ${\mathcal{O}}$ is likewise made of two $\psi$'s.

Consider the expectation value \begin{equation}
\langle \bar{\mathcal{O}}_{\bar\Omega,\bar\bP} \mathcal{O}_{\Omega,\bP}\rangle =  \frac{\displaystyle{\int\left[\prod_{\Lambda}D\bar\psi D\psi\right]}e^{-S_{\Lambda}[\bar\psi,\psi]} \bar{\mathcal{O}}_{\bar\Omega,\bar\bP} \mathcal{O}_{\Omega,\bP}}{\displaystyle{\int\left[\prod_{\Lambda}D\bar\psi D\psi\right]}e^{-S_{\Lambda}[\bar\psi,\psi]}}
\end{equation}

 In the first stage of the RG transformation, we must integrate out all fields at momenta between $\Lambda/s $ and $\Lambda$, so that the remaining action only has terms that are at momenta $<\Lambda/s$. This is a complicated process for the composite operator  since it contains terms that must be integrated out. We can write the composite operator as

 \begin{eqnarray}
 \mathcal{O}_{\Omega,\bP} =   \mathcal{O}^{<,<}_{\Omega,\bP} + \mathcal{O}^{<,>}_{\Omega,\bP}+ \mathcal{O}^{>,<}_{\Omega,\bP}+ \mathcal{O}^{>,>}_{\Omega,\bP}
 \end{eqnarray}
 where the $<,>$ denote whether the two fields entering the operator are below or above the cutoff respectively. The operator $\mathcal{O}^{<,<}_{\Omega,\bP} $ is the descendant of the composite operator in the theory with cutoff $\Lambda/s$; $\mathcal{O}^{>,<}_{\Omega,\bP} $ and $\mathcal{O}^{<,>}_{\Omega,\bP} $ are mixed terms, whose $>$ field must be integrated over; and $\mathcal{O}^{>,>}_{\Omega,\bP} $ is composite operator made up entirely of fast modes that will be integrated out.
In integrating over the fast modes we may  functionally  average over the free field action,  since the deviations from the free action will only produces terms of higher order in $g$.

 When we take the product of composite operators, we note that only terms with even numbers of `fast' ($>$) modes will survive the functional average over fast modes. Of these, one is the term $\bar{\mathcal{O}}^{<,<}_{\bar\Omega,\bar\bP}\mathcal{O}^{<,<}_{\Omega,\bP}$, which is the descendant of the two-point function in the lowered-cutoff theory; then there are the mixed terms \footnote{The other mixed possibilities are forbidden because only like spins may be contracted.}, $\bar{\mathcal{O}}^{<,>}_{\bar\Omega,\bar\bP}\mathcal{O}^{>,<}_{\Omega,\bP}$ and $\bar{\mathcal{O}}^{>,<}_{\bar\Omega,\bar\bP}\mathcal{O}^{<,>}_{\Omega,\bP}$; and finally there is the piece made up entirely of fast modes, $\bar{\mathcal{O}}^{>,>}_{\bar\Omega,\bar\bP}\mathcal{O}^{>,>}_{\Omega,\bP}$. The first term is analogous to $\psi^<\psi^<$ in our earlier discussion of the fermion Green's function. The remaining three pieces have no analog for non-composite operators, and we must determine how they alter the RG flow.

 They clearly form additive corrections to what comes from the descendant term. {\em Just for computing this additive term} we simplify things  by invoking a familiar result from computing the $T=0$ Cooper bubble explicitly: namely, that the correlations  depend on $(\bP,\Omega)$
only via the combination $\tilde{\Omega}=\sqrt{\Omega^2 + P^2}$. So we will simplify the following  discussion by choosing $\bP=0$.

Now there are no mixed terms since $|\bK + {\bP \over 2}|= |-\bK + {\bP \over 2}| $ (when $\bP=0$)  and both fields in  $\bar{\mathcal{O}}_{{\bar\Omega},0}$ and $\mathcal{O}^{}_{\Omega,0}$ are either above the cut-off or below it.

This leaves only the term consisting entirely of fast modes.  It remains  to integrate over $\mathbf{K},\omega$, i.e. we have (on converting to the Fermi-surface coordinates, performing the trivial angular integral, and recalling that all stray factors of $K_F$ from the measure are absorbed into the definitions of the fields)

\begin{eqnarray}\label{inhomog}
\langle\bar{\mathcal{O}}^{>,>}_{\Omega,0}\mathcal{O}^{>,>}_{\Omega,0}\rangle  &\rightarrow& 2\times \int_{\Lambda/s}^\Lambda\frac{d k}{2\pi}\int_{-\infty}^{\infty} \frac{d\omega}{2\pi} \frac{1}{\omega^2 +v_F^2k^2}\nonumber\\ &\sim&\frac{1}{2\pi v_F} \left(1-\frac{1}{s}\right)
\end{eqnarray}

As was mentioned earlier, this is a term of order $-d\Lambda/\Lambda  = 1-1/s=d\sigma$; we show below that this leads to an inhomogeneous term in the C-S equation for the two-point function. As an aside, we point out (but do not prove) that for any higher correlation function of composite operators beyond the two-point function, and for any other correlation functions with external legs and composite operator insertions, this contribution is of a higher order in $d\sigma $, and will not lead to inhomogeneous terms. For aficionados of the RG for the $\phi^4$ theory, this is the fermionic analog of the statement that the specific heat is the only operator that is not multiplicatively renormalizable, the signature of which is the appearance of inhomogeneous terms in the Callan-Symanzik equation for the specific heat\cite{ZinnJustin}.

Finally, we turn to the more familiar term, which is the `descendant' operator in the cutoff theory. We still need to determine how this behaves under the next two RG steps. It is sufficient to consider how the composite operator gets rescaled; the two-point function scaling follows immediately.
Before implementing the RG transformation, we should rewrite the composite operator in terms of the momentum relative to the Fermi momentum. Using the fact that  $\mathbf{K} = (K_F +k) \hat{\mathbf{K}}$, where $k \in[-\Lambda, \Lambda]$ at any stage of the RG,  on expanding to linear order in the momentum $P$, we find  the displacement from the Fermi surface is given by $ k_{\pm} \equiv \left|\pm\mathbf{K} + \frac{\mathbf{P}}{2}\right| - K_F  \approx k \pm P\cos\theta_{kP} $, while the angular coordinates are given by $\theta_{+} = \theta_k - \frac{P}{2K_F}\sin\theta_{kP},
\theta_{-} = \pi+ \theta_k + \frac{P}{2K_F}\sin\theta_{kP}$. Therefore,
\begin{eqnarray}
\bar{\mathcal{O}}_{\Omega,\bP}
&=&\int^{\Lambda}_{k,\omega}  \, \bar\psi_\uparrow\left(\omega+\Omega,  k + P\cos\theta_{kq} , \theta_k - \frac{P}{2K_F}\sin\theta_{kP}\right) \nonumber
\\& \times& \bar\psi_{\downarrow}\left(-\omega,  k - P\cos\theta_{kP}, \pi+ \theta_k + \frac{P}{2K_F}\sin\theta_{kP}  \right) \nonumber\\&\times& \Theta(\Lambda - |k_{+}|)\Theta(\Lambda-|k_-|)\nonumber
 \end{eqnarray}
 where $\int^{\Lambda}_{k,\omega} \equiv \int_{-\infty}^{\infty} \frac{d\omega}{2\pi}\int_0^{2\pi} \frac{d\theta_k}{2\pi} \int_{-\Lambda}^{\Lambda} \frac{dk}{2\pi} $. We have absorbed a factor of $\sqrt{K_F}$ into the definition of the fields, as is usual in the RG. The last two factors are to ensure that the actual momenta in the fermion lines are consistently within the cutoff.

It follows that the descendant operator is
\begin{eqnarray}
\bar{\mathcal{O}}^{<,<}_{\Omega,\bP}
&=&\int^{\Lambda/s}_{k,\omega}  \, \bar\psi^<_\uparrow\left(\omega+\Omega,  k + P\cos\theta_{kP} , \theta_k - \frac{P}{2K_F}\sin\theta_{kP}\right) \nonumber
\\& \times& \bar\psi^<_{\downarrow}\left(-\omega,  k - P\cos\theta_{kP}, \pi+ \theta_k + \frac{P}{2K_F}\sin\theta_{kP}  \right) \nonumber
 \end{eqnarray}
where we have used the step functions\footnote{Note that if both $k_{\pm}$ lie within the cutoff, $k$ itself must lie within the cutoff.}
to set the cutoff to $\Lambda/s$. In this expression, we should now redefine $(k' = sk, \ \omega' = s\omega)$ to shift the cutoff back to its full value, and then use the definition of the field rescaling once again, to find\footnote{Note that the RG does not change angles, a fact that is used implicitly here.}
\begin{widetext}
\begin{eqnarray}
\bar{\mathcal{O}}^{<,<}_{\Omega,\bP}
&=&s^{-2}\tilde{Z}_{\psi}\int^{\Lambda}_{k',\omega'}  \, \bar\psi'_\uparrow\left(\omega'+s\Omega,  k' + sP\cos\theta_{k'P} , \theta_{k'} - \frac{P\sin\theta_{k'P}}{2K_F}\right) \times \bar\psi'_{\downarrow}\left(-\omega',  k' - sP\cos\theta_{k'P}, \pi+ \theta_{k'} + \frac{P\sin\theta_{k'P}}{2K_F}  \right) \nonumber
 \end{eqnarray}
\end{widetext}

This is the tree-level rescaling of the composite operator; at one loop, the composite operator acquires an anomalous dimension, whose effects may be captured with another renormalization parameter $\tilde{Z}_{\mathcal{O}}$ (we lump the factor of $s^{-2}$ with this as well). Thus, we have the overall rescaling:

\begin{equation}\label{compscaling}
\bar{\mathcal{O}}^{<,<}_{\Omega,\bP} =  \tilde{Z}_{\mathcal{O}}^{1/2} \tilde{Z}_{\psi} \bar{\mathcal{O}}'_{s\Omega,s\bP}
\end{equation}
where the $'$ denotes the fact that we have related fields in two different theories, the original and the one after the RG step.

Following the same steps used in deriving the C-S equation previously, and combining the scaling of the descendant operator (Eq.~\ref{compscaling})  with the inhomogeneous term (Eq.~\ref{inhomog}) contributed by the pieces of the composite operator lying above the cutoff, we have the following equation obeyed by the flow of the two-point function $\langle\bar{\mathcal{O}}\mathcal{O}\rangle$:

\begin{eqnarray}
\lefteqn{\langle \bar{\mathcal{O}}_{\bar\Omega,\bar\bP} \mathcal{O}_{\Omega,\bP}\rangle_{S,\Lambda} }\\ & &= \frac{1}{2\pi v_F} \left(1-\frac{1}{s}\right) + \tilde{Z}_{\mathcal{O}}\tilde{Z}_{\psi}^2 \langle \bar{\mathcal{O}}_{s\bar\Omega,s\bar\bP} \mathcal{O}_{s\Omega,s\bP}\rangle_{S_s,\Lambda}
\end{eqnarray}

Once again, we may use the fact that the left hand side and thus the right hand side of this expression are independent of $s$ to perform the same manipulations as before, and arrive at the expression
\begin{eqnarray}
0 &=&\frac{1}{2\pi v_F} + \left[{\beta}(g) \frac{\partial}{\partial g}+ 2+2{\gamma}_{\mathcal{O}}\right] \langle \bar{\mathcal{O}}_{\bar\Omega,\bar\bP} \mathcal{O}_{\Omega,\bP}\rangle_{S,\Lambda} \nonumber\\ & &+\left\langle\left.\frac{d}{ds}\right|_{s\rightarrow1} \bar{\mathcal{O}}_{s\bar\Omega,s\bar\bP} \mathcal{O}_{s\Omega,s\bP}\right\rangle_{S,\Lambda}
\end{eqnarray}
where we have defined $ 2+2{\gamma}_{\mathcal{O}} =\left. \frac{d\log\left(\tilde{Z}_{\mathcal{O}}\tilde{Z}_{\psi}^2\right)}{ds}\right|_{s\rightarrow 1} $

We leave it as an exercise to the reader to show that the final term on the right hand side is equivalent to the action of the operator $\lder{\Omega} +\lder{P}$ on the two point correlation function; any discrepancies between the two vanish because of the symmetry of the Fermi surface. Making this substitution, we find

 \begin{eqnarray}
 & &\left[\lder{\Omega}+\lder{P} +{\beta}(g) \frac{\partial}{\partial g}+ 2+2{\gamma}_{\mathcal{O}}\right]\ \langle \bar{\mathcal{O}}_{\bar\Omega,\bar\bP} \mathcal{O}_{\Omega,\bP}\rangle_{S,\Lambda} \nonumber \\&=& -\frac{1}{2\pi v_F}
\end{eqnarray}

As a final step, we note that once again we need to remove a trivial delta function in going between the expectation value and the correlation function,
$\langle \bar{\mathcal{O}}_{\bar\Omega,\bar\bP} \mathcal{O}_{\Omega,\bP}\rangle_{S,\Lambda} \equiv \Gamma_{\mathcal{O}}^{(0,2)}(\Omega, q) \delta_{\bar\Omega,\Omega} \delta_{\bar\bP,\bP}$. Thus, we finally arrive at the inhomogeneous Callan-Symanzik equation for the two-point Cooper pair correlator,
\begin{widetext}
 \begin{equation}
 \left[\lder{\Omega}+\lder{P} +{\beta}(g) \frac{\partial}{\partial g}+2{\gamma}_{\mathcal{O}}\right] \Gamma_{\mathcal{O}}^{(0,2)}(\Omega,q;g,\Lambda)=  -\frac{1}{2\pi v_F}
\end{equation}
\end{widetext}

\section{Solution of the Cooper Pair Callan-Symanzik Equation}
The next step is to solve the Callan-Symanzik equations: given a bare coupling constant $g_0$, we wish to determine the long-wavelength, low-frequency behavior of $\Gamma_{\mathcal{O}}^{(0,2)}$. Since the dimensionless combination of the frequency, momentum, and cutoff that enter the correlation functions must be of the form $\sqrt{\Omega^2 + P^2}/\Lambda \equiv \tilde{\Omega}/\Lambda$, (where a phenomenological velocity - that depends on the Fermi-liquid parameters such as $m^*$ - has been set equal to unity.\footnote{Such phenomenological terms cannot in general be obtained within the RG.})  Using this, we argue that $\lder{\Omega}+\lder{P} \leftrightarrow -\lder{\Lambda}$ when acting on the correlation functions; this gives the slightly more tractable equation

 \begin{equation}
 \left[\lder{\Lambda}- {\beta}(g) \frac{\partial}{\partial g}-2{\gamma}_{\mathcal{O}}\right] \Gamma_{\mathcal{O}}^{(0,2)}(\tilde{\Omega};g,\Lambda)=  \frac{1}{2\pi v_F}
\end{equation}

We may  now solve this  equation by the method of characteristics \cite{ZinnJustin}, and find
\begin{widetext}
\begin{eqnarray}
\Gamma^{(0,2)}_{\mathcal{O}}(\tilde{\Omega};g_0,\Lambda) &=& e^{-2\int_{1}^{\frac{\tilde{\Omega}}{\Lambda}}\frac{dx}{x}\,{\gamma_{\mathcal{O}}(g(x))}}\Gamma^{(0,2)}_{\mathcal{O}}\left(1;g\left(\frac{\tilde{\Omega}}{\Lambda}\right),1\right)
- \frac{1}{2\pi v_F} \int_{1}^{\frac{\tilde{\Omega}}{\Lambda}}\frac{dx}{x}\,  e^{-2\int_{1}^{x}\frac{dy}{y}\,{\gamma_{\mathcal{O}}(g(y))}}
 \nonumber \\
\text{with}\,\,\,\,\, x\frac{d}{dx} g(x) &=& -\beta(g(x))\,\,\,\, \text{and} \,\,\,\, g(x=0) \equiv g_0
\end{eqnarray}
\end{widetext}

Before we can complete our solution, we need to compute $\beta$ and $\gamma_{\mathcal{O}}$. We sketch the calculation in the Appendix, and just cite the results here:
\begin{eqnarray}
\beta(g) &=& -a g^2 \nonumber\\
\gamma_{\mathcal{O}}(g) &=& - ag
\end{eqnarray}
with $a>0$.

With these in hand, we find that
\begin{eqnarray}\label{CSsol}
\Gamma^{(0,2)}_{\mathcal{O}}(\tilde{\Omega};g_0,\Lambda) &=& \left[\frac{g\left(\frac{\tilde{\Omega}}{\Lambda}\right)}{g_0} \right]^2 \Gamma^{(0,2)}_{\mathcal{O}}\left(1;g\left(\frac{\tilde{\Omega}}{\Lambda}\right),1\right) \nonumber \\ & &- \frac{1}{2\pi v_F g_0^2} \left[ g\left(\frac{\tilde{\Omega}}{\Lambda}\right) - g_0\right]
\end{eqnarray}

where the flow of the coupling is given by
\begin{equation}
g\left(\frac{\tilde{\Omega}}{\Lambda}\right)  = \frac{g_0}{1 - a g_0 \log\frac{\tilde{\Omega}}{\Lambda}}
\end{equation}
which clearly reflects the fact that $g$ is marginal: as we take $\tilde{\Omega}\rightarrow 0$, $g$ vanishes logarithmically. Note that we must always have $\tilde{\Omega}<\Lambda$, so that the logarithm in the denominator is positive and does not lead to any singularity as we take $\tilde{\Omega}\rightarrow 0$.

Since the fact that $g\left(\frac{\tilde{\Omega}}{\Lambda}\right)$ is marginal, in the limit of interest, the second term in Eq.~\ref{CSsol} dominates, and we have
\begin{eqnarray}\label{CSsol2}
\Gamma^{(0,2)}_{\mathcal{O}}(\tilde{\Omega};g,\Lambda) &\sim& - \frac{1}{2\pi v_F g_0^2} \left[ g\left(\frac{\tilde{\Omega}}{\Lambda}\right) - g_0\right]  \nonumber\\ &=& \frac{1}{2\pi v_F} \frac{a \left|\log\frac{\tilde{\Omega}}{\Lambda}\right|}{1 + ag_0  \left|\log\frac{\tilde{\Omega}}{\Lambda}\right|}
\end{eqnarray}

We see that for $g_0$ strictly zero, the expression diverges logarithmically as $\tilde{\Omega}\rightarrow 0$, reflecting the singularity in the zero-frequency, zero-momentum pairing response of the free Fermi gas at $T=0$. However, for any finite $g_0$, we find that the response is nondivergent:
$ \Gamma^{(0,2)}_{\mathcal{O}}(\tilde{\Omega};g,\Lambda)  \sim 1/2\pi v_Fg_0$ as $\tilde{\Omega} \to 0$. Observe that,
nicely enough, this answer itself diverges as $g_0 \rightarrow 0$.

\section{Another route}

We have outlined our derivation above at some length for we were interested in a particular method of getting the answer,
in which we follow the irrelevant coupling all the way to zero while continuing to renormalize. For the generic marginal
coupling, this is the easiest way to go and even for fermions there are other problems, e.g. involving gauge fields, where we expect this technique will be the way to go.

However, there is another route to our answer---as readers may guess by looking at it. In this approach we renormalize until
we get to an exactly solvable problem and then we appeal to the exact results. In our problem, the action with the purely
BCS channel interaction corresponds to the reduced BCS Hamiltonian, which has infinite range interactions and is
thereforw exactly solvable by a saddle point computation in the infinite volume limit. The same method shows that the RPA result
for the superconducting susceptibility for this problem is exact. This has precisely the form (\ref{CSsol2}) with $\Lambda$ now
being the scale at which we switch to the exact solution. Adding in the additional operator renormalizations gathered {\it en
route} will change the answer but not the finiteness of the result or its behavior as $g_0 \rightarrow 0$. Indeed, in this
approach it is also straightforward to explicitly include the Landau couplings as the resulting Hamiltonian is still
exactly solvable \cite{Lehmann}.

The general procedure we have described in this section is also what is used in the implementations of the the RG for
interacting fermions known as the Functional Renormalization Group (FRG), see for example Appendix B in Ref.~\onlinecite{Honerkamp01}.
The difference is that as in such work generally relevant flows with multiple coupling constants get stopped at some scale the
resulting problem is not typically exactly solvable in a controlled sense. However, that has to do with the ends to which the
FRG is put---the idea is the same.

\section{Discussion}

Our result (\ref{CSsol2}) verifies the claim with which we began---namely that the superconducting susceptibility of the
Fermi liquid is finite due to the intercession of the marginally irrelevant BCS coupling that is present even for repulsive
interactions. This exact compensation of the leading singularity by the irrelevant flow is somewhat surprising from the RG
perspective, certainly if you compare with the results on ferromagnets in four dimensions that we reviewed in the introduction.
Possibly multi-band systems will lead to richer possibilities but that is a subject for future work. It is also of interest
to extend the RG approach taken here to the derivation of the effects of the marginal flow on the electron Green's function.
As this requires application of the RG at two loops, this will require going beyond the straightforward momentum shell method
used in this paper.

\begin{acknowledgements}
We would like to thank R. Thomale for useful comments on the manuscript, and in particular for clarifying the relation of the work presented here to the FRG. We also acknowledge useful discussions with B.A. Bernevig. 
\end{acknowledgements}
\begin{appendix}

\section{Calculation of the Beta Function and Anomalous Dimension}
While we expect many of our readers to be familiar with the computation of the RG functions, for the sake of completeness, and because the anomalous dimension calculation may be unfamiliar to some, we briefly review the procedure here. Following
Shankar \cite{RSRMP}, we compute corrections to the action from integrating out modes between $\Lambda-d\Lambda$ to $\Lambda$ via the cumulant expansion. Essentially, we treat the coupling as a perturbation $\delta S$, and use the result\begin{equation}
\langle e^{\delta S} \rangle  = e^{\left[\langle \delta S \rangle  +\frac{1}{2}\langle (\delta S)^2 \rangle -  \left(\langle\delta S\rangle\right)^2 \right]}
\end{equation}

\begin{figure}
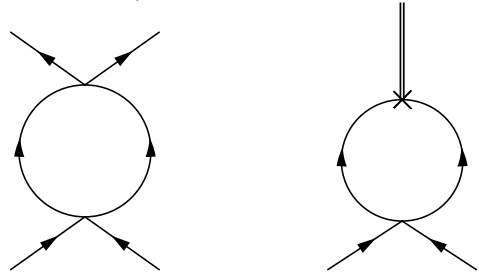

{\includegraphics[width=2cm]{OneLoopBeta.1} \hspace{2cm}
\includegraphics[width=2cm]{OneLoopAnomDim.1}}
\caption{\label{OneLoopDiagrams}One-loop diagrams for the beta function and anomalous dimension of $\bar{\psi}\bar{\psi}$}
\end{figure}

where all averages are over the modes being integrated out. In each term, we are to select the number of `external' legs, which will belong to the modes below the shell of integration, and  this determines which term in the action will be corrected by the term obtained by integrating out the remaining  fields. In order to determine the anomalous dimension of the Cooper pair operator, the easiest method is to add a source term of the form $\mathcal{J}_\mathcal{O}\psi\psi$ (and its complex conjugate) to the action and determine how it gets renormalized at one loop; a moment's thought will suffice to realize that this is equivalent to determining $\gamma_{\mathcal{O}}$.

The one-loop diagrams contributing to the $\beta$-function and the anomalous dimension of $\mathcal{O} $ are shown in Fig. \ref{OneLoopDiagrams}.

Evaluating the first diagram, we find that $\delta g =  a g^2 \frac{d\Lambda}{\Lambda}$, or in other words that $\beta(g) = \left.\frac{dg}{ds}\right|_{s\to 0} = -a g^2$, since $ds = -d\Lambda/\Lambda$ . From the second diagram, we find that $\frac{\delta \mathcal{J}_\mathcal{O}}{\mathcal{J}_\mathcal{O}} = a g \frac{d\Lambda}{\Lambda}$; combining this with the tree level rescaling we find that $\mathcal{J}_\mathcal{O} \rightarrow s^{1-ag} \mathcal{J}_\mathcal{O}$, which gives us the result (cf. Eq.~\ref{compscaling}) that $2+2\gamma_{\mathcal{O}} =  2-2ag$, or in other words, that $\gamma_\mathcal{O} = -ag$. In these expressions, $a$ is a positive constant, whose value is unimportant; the significant point is that it is the {\it same} constant in both $\beta$ and $\gamma_{\mathcal{O}}$.

\end{appendix}

\bibliography{RGEFL2}

\end{document}